\documentclass[sigconf]{acmart}

\AtBeginDocument{%
  \providecommand\BibTeX{{%
    \normalfont B\kern-0.5em{\scshape i\kern-0.25em b}\kern-0.8em\TeX}}}

\setcopyright{acmcopyright}
\copyrightyear{2019}
\acmYear{2019}
\acmDOI{XX.XXXX/XXXX}

\acmConference[IUI '20]{IUI '20: ACM International Conference on Intelligent User Interfaces}{March 17--20, 2020}{Cagliari, Italy}
\acmPrice{15.00}
\acmISBN{978-1-XXXXX}

\usepackage{balance}       
\usepackage{graphics}      
\usepackage[T1]{fontenc}   
\usepackage{txfonts}
\usepackage{mathptmx}
\usepackage{color}
\usepackage{booktabs}
\usepackage{textcomp}
\usepackage{threeparttable}
\usepackage{multirow}
\usepackage[noend]{algorithm2e}
\usepackage{microtype}        
\usepackage{hyperref}
\usepackage{ccicons}          
\usepackage{todonotes}

\renewcommand{\textrightarrow}{$\rightarrow$}

\begin{document}

\title{VASTA: A Vision and Language-assisted Smartphone Task Automation System}

\author{Alborz Rezazadeh Sereshkeh, Gary Leung, Krish Perumal, Caleb Phillips, Minfan Zhang, Afsaneh Fazly, Iqbal Mohomed}
\affiliation{%
  \institution{Samsung AI Research Centre}
  \city{Toronto}
  \country{Canada}}


\begin{abstract}
We present VASTA, a novel vision and language-assisted Programming By Demonstration (PBD) system for smartphone task automation. Development of a robust PBD automation system requires overcoming three key challenges: first, how to make a particular demonstration robust to positional and visual changes in the user interface (UI) elements; secondly, how to recognize changes in the automation parameters to make the demonstration as generalizable as possible; and thirdly, how to recognize from the user utterance what automation the user wishes to carry out. To address the first challenge, VASTA leverages state-of-the-art computer vision techniques, including object detection and optical character recognition, to accurately label interactions demonstrated by a user, without relying on the underlying UI structures. To address the second and third challenges, VASTA takes advantage of advanced natural language understanding algorithms for analyzing the user utterance to trigger the VASTA automation scripts, and to determine the automation parameters for generalization. We run an initial user study that demonstrates the effectiveness of VASTA at clustering user utterances, understanding changes in the automation parameters, detecting desired UI elements, and, most importantly, automating various tasks. A demo video of the system is available here: \url{http://y2u.be/kr2xE-FixjI}.
\end{abstract}




\maketitle

\section{Introduction} \label{sec:intro}

Today's smartphones provide a sophisticated set of tools and applications that allow users to perform many complex tasks. Given the diversity of existing tasks and the ever-increasing amount of time users spend on their phones, automating the most tedious and repetitive tasks (such as ordering a pizza or checking one's grades using a school app) is a desirable goal for smartphone manufacturers and users alike.

\par Intelligent assistants such as Alexa, Siri, Bixby, and Google can be used to automate and voice-enable particular tasks such as web search and device control. However, the functionality of such agents is usually limited to built-in smartphone apps (message, calendar, etc.) and a small number of integrated web services and external apps. In other words, they are unable to control most third-party apps due to the significant variations in apps and tasks. 

\par An emerging method for enabling users to automate smartphone tasks is Programming-by-Demonstration (PBD). This method enables smartphone users to automate their activities by simply performing them once or multiple times, which makes it appealing for smartphone users with little or no programming knowledge. 


\par Given the ever-evolving landscape of applications and tools available, a smartphone task-automation solution must ideally overcome the following challenges: (i) being agnostic of the apps, user inputs (taps, swipes, ...), and User Interface (UI) elements involved in the task, (ii) being robust to positional (e.g., shift in the location of UI elements) and visual changes (e.g., changes caused by updates) in apps, (iii) being able to recognize from the user utterance what automation the user wishes to carry out (e.g., a new automation script or one that has been previously learned), and (iv) being able to recognize changes in the automation parameters using the user utterance (e.g., the pizza type in a pizza ordering task).

\par In this paper, in attempt to address all the challenges above, we introduce VASTA, a PBD-based smartphone task automation system, which enables users to automate a wide variety of tasks regardless of the applications involved. VASTA overcomes the first two challenges by taking advantage of state-of-the-art computer vision algorithms, including object detection and optical character recognition (OCR), to accurately label UI elements without relying on the underlying UI structures. To address the last two challenges, VASTA leverages natural language understanding (NLU) algorithms to process the user utterance and detect its parameters for generalization.

\subsection{Contributions}
To summarize, the contributions of this paper are:

\textbf{(I)} VASTA, a PBD system that leverages state-of-the-art vision and language algorithms to enable smartphone users to create and execute an automation script for arbitrary tasks using any or multiple third-party apps. Moreover, the demonstration is exactly the sequence of actions that the user would perform in the course of their task (not requiring any special annotations or scripting by the user), thus making the approach more accessible to non-expert end-users. Relying only on capturing the screen's state and <X, Y> coordinates of user interaction, this system can provide intelligent vision-based automation to any application, and potentially platform, where these fields are available. 

\textbf{(II)} A natural language understanding component to automatically cluster user utterances belonging to the same 
task, and predict their parameters.

\textbf{(III)} An object detector network for detecting UI elements based on a state-of-the-art object detection model trained on a large dataset of UI elements. 

\textbf{(IV)} User studies to evaluate the performance of different modules of VASTA as well as the end-to-end system.

\section{Related Work} \label{sec:relwork}
\subsection{Smartphone Task Automation Using PBD}

\par There have been several attempts in using PBD for task automation on smartphones. \textit{Keep-doing-it} \cite{maues2013keep} utilizes a user's demonstration to derive automation rules, expressed in terms of Event-Condition-Action \cite{aztiria2008autonomous}. However, this approach has limited applicability and can only be useful in automating short sequences of actions, such as WiFi toggle, and cannot be used to control arbitrary third-party apps.

\par Another approach for smartphone task automation using PBD is to leverage macro-recording tools on smartphones \cite{rodrigues2015breaking}. Using such tools enables users to automate a longer sequence of actions. However, only the exact demonstrated procedure can be replayed. Also, such systems fail when there are changes in the UI.

\par Another approach that has recently gained attention, is to leverage the accessibility API provided by the smartphone's operating system (OS). \textsc{Sugilite} \cite{SUGILITE} is an example of such a system, which takes advantage of the app's UI hierarchy structure, provided by Android accessibility API, to create an automation script. Nonetheless, due to the limitations of accessibility mechanisms, \textsc{Sugilite} cannot be used in some automation scenarios, such as interactions with unlabeled graphical icons.

\par In this paper, we choose a different approach from all previous smartphone PBD automation systems and rely on computer vision techniques to correctly label demonstrated interactions using the captured screen-state and <X, Y> coordinates of user interaction. To the best of our knowledge, this is the first attempt to rely on computer vision techniques for the development of a task automation system for smartphones. 

\subsection{Utilizing Computer Vision Algorithms in the UI Domain}

\par The first attempt of analyzing the visual patterns rendered on the screen to support interactions was reported in the late 90s \cite{potter1999pixel}, in which researchers investigated the potential of \textit{Direct-Pixel-Access} in supporting application-independent end-user programming. A decade later, Yeh et al. \cite{robert2009sikuli} presented Sikuli, a visual approach to search and automate graphical user interfaces (GUI) using screenshots. Sikuli introduces a help system which allows users to query about a GUI element (e.g., an icon or a toolbar button) by taking its screenshot. Also, it leverages screenshot patterns to provide a visual scripting API for automating GUI interactions. However, creating an automation script with this approach requires a large amount of programming by the users and hence, is successful mostly for simple tasks. Intharah et al. \cite{intharah2017help} address this challenge by creating a computer vision-based PBD system, called HILC. However, the user is still required to answer follow-up questions after the demonstration to disambiguate similar items. Also, instead of learning appearance features, the system relies on fixed appearance models with fixed size and aspect ratio which, for instance, fail when items in a list are short and wide \cite{intharah2017help}. 

In this work, with the help of recent large datasets of UI elements such as ERICA \cite{deka2016erica} and RICO \cite{deka2017rico}, we are able to leverage state-of-the-art deep learning techniques in object detection to develop a PBD system which is not limited to any fixed appearance model. Instead, it learns the appearance model of UI elements from data (without extra work by users) and can thus generalize to novel items.

\subsection{Task Automation using Natural Language Processing}
There has been significant previous work (such as Almond \cite{Almond2} and SUGILITE \cite{SUGILITE}) on automating personal assistant tasks using natural language processing. These require significant training data in the form of task-specific datasets of natural language commands mapped to instructions.

\par There is also existing work on employing minimal developer efforts to bootstrap the automation of personal assistant tasks. UIVoice enables the development of third-party voice user interfaces on top of existing mobile applications \cite{tarakji2018voice}. It requires a user/developer to define the intents, parameters and sample utterances for a new task. The CRUISE \cite{shen2018cruise} system uses rule-based and data-driven algorithms to iteratively generate utterances from a few seed utterances/phrases or an intent verb phrase defined by the developer. The developer is also responsible for pruning incorrectly generated utterances. KITE uses hand-crafted semantic rules based on UI elements to detect tasks and their parameters \cite{li2018kite}. It uses a neural network transduction model to prompt the user for missing parameters, in addition to developer effort in manually revising bot templates, correcting errors, etc. SUGILITE's \cite{SUGILITE} conversational agent employs a Learning by Instruction Agent (LIA) \cite{azaria2016instructable} that parses verbal commands with a Combinatory Categorical Grammar (CCG) parser \cite{steedman2009combinatory}, which requires hand-engineering of lexicalized rules. PUMICE \cite{li2019pumice} builds on top of SUGILITE to handle tasks with conditional structure using the SEMPRE framework \cite{berant2013semantic} that requires less training data than other neural network-based approaches. There also exists relevant work on semantic parsing \cite{zettlemoyer2005ccg}, frame-semantic parsing \cite{swayamdipta2017frame}, and dependency-based parsing \cite{reddy2017dep}, but all of these require large training data specific to the task or domain of interest.

In contrast to previous work mentioned above, VASTA's language component comprises of an unsupervised method (that does not require manually labeled data) to detect task intent, and a method that requires only one labeled utterance (and no manual engineering of lexical or semantic rules) to detect task parameters. Our system's language component is related to an existing line of work that can automate tasks using labeled training data from other task domains. One work employs a neural network model trained on abundant labeled data in a different task domain along with embeddings of the task parameters \cite{bapna2017towards}. Another work \cite{hakkani2015clustering} clusters novel intents by leveraging similarities in the semantic parse trees of utterances. The former is limited by its heavy reliance on commonalities in parameters of different tasks, and the latter is limited by the requirement of a large dataset of unlabeled utterances.

\section{System Overview} \label{sec:system}
\begin{figure}[h!]
\centering
\includegraphics[width=0.48\textwidth]{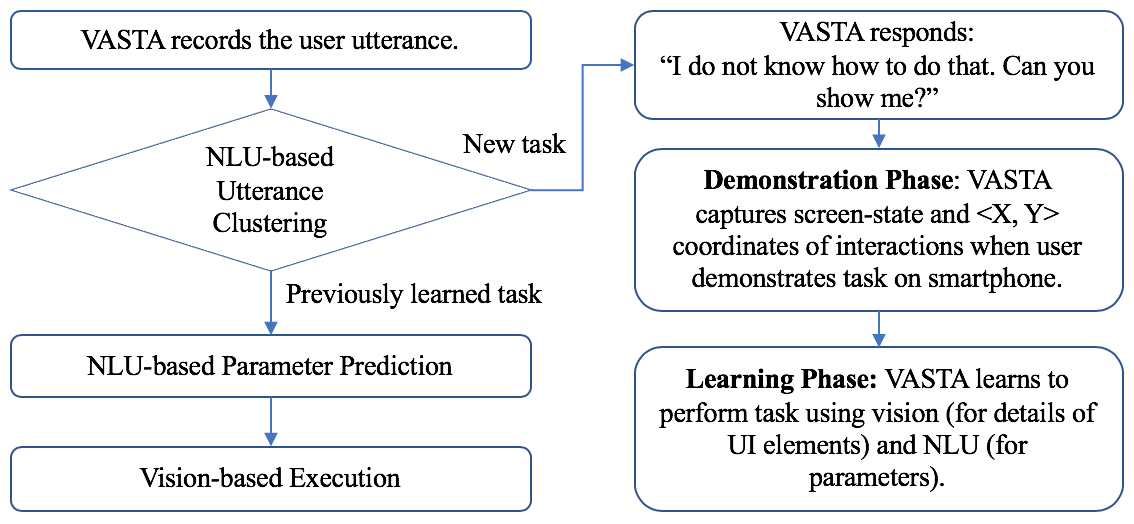}
\caption{An overview of the VASTA system.} 
\label{fig:overview}
\end{figure}

Figure \ref{fig:overview} presents an overview of the VASTA system. To automate a task, first, the user needs to give a voice command (which, henceforth, we will call "voice utterance") to VASTA. The voice utterance is converted to text using Google Cloud Speech-to-Text. VASTA analyzes the text utterance using NLU to determine if it refers to a new task or an existing one for which a demonstration is already provided by the user. If it is a new task, VASTA replies: "I do not know how to do that. Can you show me?". If the user answers yes, the demonstration phase starts. The user performs the task (s)he would like to create an automation for and then shakes the phone to end the demonstration.

\par After the end of the demonstration phase, VASTA enters the learning phase, during which it utilizes an object detection network\footnote{Object detection is a computer vision and image processing technique that deals with detecting the presence and locating the instances of objects in an image.} to recognize the bounding boxes\footnote{Each box around an instance of an object is called a bounding box.} of the UI elements that the user interacted with at each step of the demonstration. By analyzing the utterance as well as the demonstration's interaction trace (e.g., the message that the user typed or the textual information written on the UI elements that the user clicked on), VASTA creates and parameterizes a script. 

\par After the learning phase, the user can execute the task using the same utterance. Also, by leveraging NLU, VASTA is able to detect user utterances that are likely to be variations of the same utterance. During the execution, if the locations of the UI elements are different from the demonstration phase, VASTA is still able to find the target UI element using computer vision. If there are visual changes in UI elements (e.g., due to updates or theming), VASTA is still able to find target UI elements using the text written on them.  

\par VASTA is able to generalize the automation if the user repeats the utterance with a different parameter. VASTA first evaluates if this new utterance is referring to one of the previously learned automation scripts. If so, using NLU, it determines what the value of the parameter(s) is and replicates the same steps with the new parameters (e.g., changes the message that the user typed in a step).
\par We provide the implementation details of different components of VASTA in the following section.

\section {System Implementation} \label{sec:implementation}

\subsection{Language Component}
VASTA requires natural language understanding to determine if an input utterance refers to a new task, or an existing one for which a demonstration is already provided by the user. In case an utterance is matched to an existing task, it also needs to determine the task parameters so that the vision component can correctly repeat the demonstration. Both of these requirements are fulfilled by the modules described below.

\subsubsection{Utterance Clustering Module}
\par
\begin{figure}[h!]
	\centering
	\includegraphics[width=0.41\textwidth]{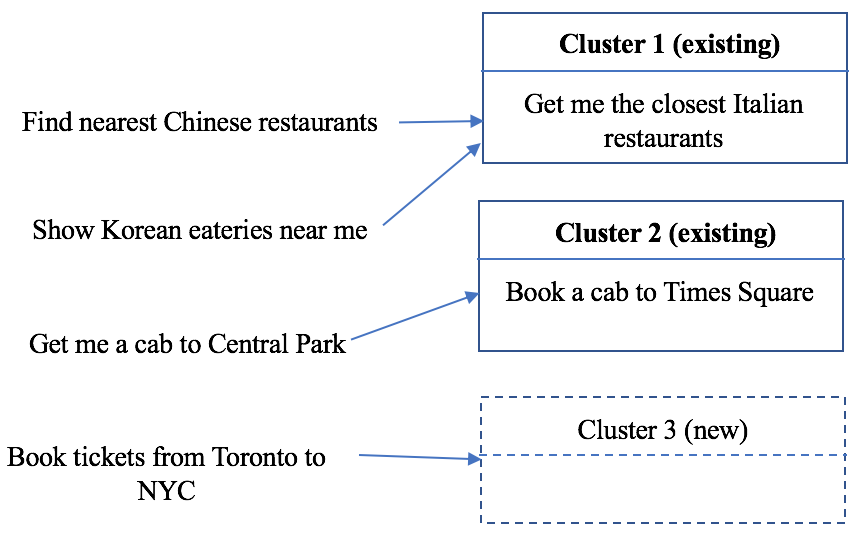}
	\caption{Example execution of Utterance Clustering Module. Each utterance on the left is either assigned to one of the existing clusters on the right or a newly created cluster.}
	\label{fig:ucm-example}
\end{figure}

 The aim of the clustering module is to determine if a new utterance refers to a new task or an existing one already trained by the user. For example, assume that a user has trained tasks for finding the nearest restaurants, using the utterance ``Get me the closest Italian restaurants'', and for booking a cab, using the utterance ``Book a cab to Times Square''. Now the user utters ``Find nearest Chinese restaurants'', which should be identified as belonging to the task of finding restaurants. Next the user utters ``Get me a cab to Central Park'', which should be matched with the task of booking a cab. Finally, if the user utters ``Book tickets from Toronto to NYC'', the utterance should be identified as belonging to a new task. This is demonstrated with an example in Figure \ref{fig:ucm-example}.

\par The Utterance Clustering Module clusters incoming user utterances that are likely to be variations of the same command (with the same or different parameters) thereby grouping together all utterances belonging to the same task. We encode each incoming utterance into a vector embedding using a pre-trained Universal Sentence Encoder \cite{cer2018encoder}, which has similar representations for semantically similar sentences. Then we compare this vector embedding with all existing cluster centroids using angular cosine similarity. The centroid of a cluster refers to the mean of the vector embeddings of all utterances in that cluster. If the similarity score is above a hard threshold (t\textsubscript{hard}), we assign the utterance to the respective cluster. Otherwise, if it is above a soft threshold (t\textsubscript{soft}), the system asks the user to verify if the utterance relates to the same task as the canonical utterance of the respective cluster. For example, for the user utterance ``Find nearest Chinese restaurants'', the system might ask the user: ``Did you mean a task similar to: `Get me the closest Italian restaurants'?''. If the user confirms, the system assigns the utterance to the cluster. Otherwise, a new cluster is created and the utterance is saved as the canonical utterance for that cluster (c\textsubscript{can}). The clustering algorithm is shown in Algorithm \ref{alg:ucm}.


\begin{algorithm}[h]
	\textbf{Inputs:} $t_{hard}$, $t_{soft}$ // hard and soft thresholds\\
	\textbf{Initialize:} $C$ = $\emptyset$ // initialize set of clusters\\
	\ForEach{u}{
		$e_u$ = encoding($u$);\\
		// best matching cluster and similarity score\\
		$bc$ = $none$, $bsim$ = $0.0$;\\ 
		\ForEach{$c \in C$}{
			$ctd$ = computeCentroid($c_{utt}$);\\
			$sim$ = computeSimilarity($e_u, ctd$);\\
			\If{$sim$ > $bsim$}{
				$bsim$ = $sim$;\\
				$bc$ = $c$;\\
			}
		}
		\If{$bsim$ > $t_{hard}$}{
			$bc_{utt}$ = $bc_{utt} \cup u$
		}
		\ElseIf{$bsim$ > $t_{soft}$}{
			// ask user if $u$ refers to the same task as $bc_{can}$\\
			askUser($bc_{can}$, $u$);\\
		}
		\Else{
		    // create new cluster $c'$ where \\
		    // $c'_{utt}$ = utterances in cluster $c'$ and \\ 
		    // $c'_{can}$ = canonical utterance of $c'$ \\
			$c'_{utt}$ = $u$, $c'_{can}$ = $u$;\\
			$C$ = $C \cup c'$;\\
		}
	}
	\vspace{0.5em}
	\caption{Utterance clustering algorithm.}
	\label{alg:ucm}
\end{algorithm}

\subsubsection{Parameter Prediction Module}
In the clustering module, if an utterance is assigned to a new task, we compare the utterance text against text in the clicked buttons or input text entered during the demonstration (similar to SUGILITE \cite{li2017sugilite}), and predict the best matching string(s) as the parameter(s). Otherwise, we pass the utterance into the Parameter Prediction Module which can predict the parameter(s) using natural language understanding. For example, when a new utterance ``Find nearest Chinese restaurants'' is added to a cluster containing the utterance ``Get me the closest Italian restaurants'' with the known parameter `Italian', we need to identify that `Chinese' is the parameter of the new utterance.

\par For predicting the parameters, we use multiple linguistic cues such as word lemmas, part-of-speech (POS) tags, word embeddings\footnote{We use pre-trained GloVe word vectors (\url{http://nlp.stanford.edu/data/glove.6B.zip}).} \cite{pennington2014glove} and dependency parse representations\footnote{We use the spaCy library (\url{https://spacy.io/}) for dependency parsing.} \cite{demarneffe2006generating}. The dependency parse of a sentence provides the grammatical relationship between each pair of words in the utterance. For example, in the utterance ``Get me the closest Italian restaurants" the dependency parse indicates an adjective modifier dependency from `Italian' to `restaurants'. Analogously, in another utterance such as ``Find nearest Chinese restaurants", the dependency parse indicates the same relationship from `Chinese' to `restaurants'. We seek to leverage this dependency similarity between words in two different utterances and match the known parameters in the canonical utterance to predict parameters in a new utterance.

\par Our parameter matching algorithm works as follows: We create a bipartite graph such that each node in the dependency parse of the canonical utterance is connected to each node in the parse of the new utterance. The edge weight is the score comprising (i) the cosine similarity between the two nodes' word embeddings, (ii) exact match of the lemmas of the nodes and their neighbours, (iii) exact match of the POS tags of the nodes and their neighbours, and (iv) exact match of the dependency labels of the two nodes' edges. Then we use a maximum weighted bipartite graph matching algorithm \cite{galil1986graph} to find the parameters in the new utterances that match with the known parameters in the canonical utterance.

\subsection{Vision Component}
The vision component of VASTA consists of three modules: object detection, template matching, and OCR. In this section, we provide a brief description of these modules. The explanation of how and where VASTA utilizes each of these modules is provided later in \ref{sec:endtoend}. 

\begin{figure}[h!]
	\centering
	\includegraphics[width=0.48\textwidth]{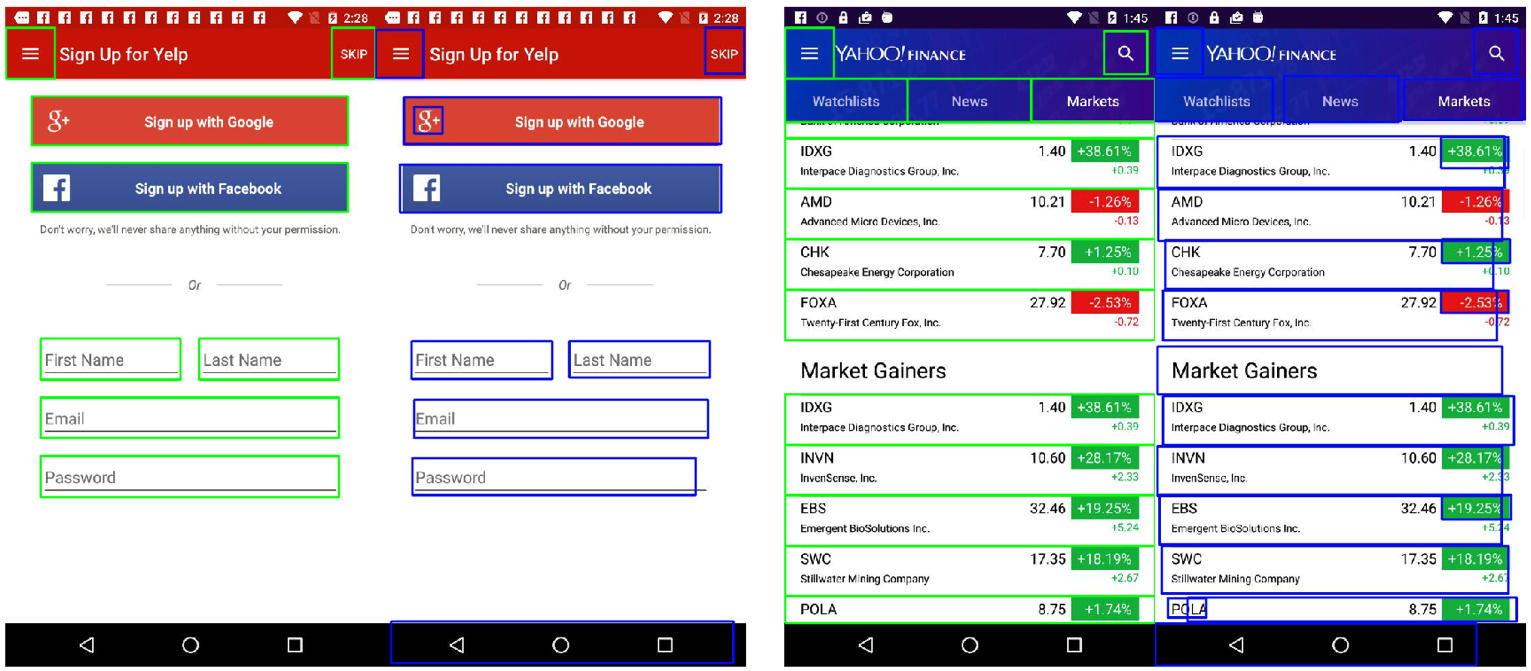}
	\caption{Two examples of UI screens provided in the RICO dataset. For each pair, the image on the left contains the bounding boxes provided by the accessibility services. The image on the right shows the bounding boxes generated by our trained object detection network based on the RetinaNet model. None of these examples were included in the training set.} 
\label{retinaNetExamples}
\end{figure}

\textbf{RetinaNet Object Detector Module:} 
In order to develop an object detector to detect UI elements on the screen, we trained a state-of-the-art object detection network, called the RetinaNet \cite{lin2018focal}, on a large dataset of UI screens (across various apps), called the RICO dataset \cite{deka2017rico}. The details of the training process, as well as the results for different object detector models that we tested, are provided in subsection \ref{subsec:expObjectDetector}. Figure \ref{retinaNetExamples} illustrates some examples of the bounding boxes generated by our trained object detection network.

\textbf{Template Matching:}
The template matching module in VASTA uses mean square difference between pixel values as a scoring function at each location. VASTA only considers exact matches for the UI element that it is searching for.

\textbf{Optical Character Recognition (OCR) module:} VASTA uses OCR to detect the text within a UI Element. We built our OCR framework using Tesseract, an open-source python library.\footnote{Tesseract OCR by Google \url{https://github.com/tesseract-ocr/}.}

\subsection{End to End System} \label{sec:endtoend}

VASTA works in three phases. First, the user needs to demonstrate to VASTA how to perform a task (Demonstration phase). Then, VASTA processes the user's actions and create a generalizable script from it (Learning phase). Finally, the next time the user asks VASTA to perform the same task, with the same or different parameters, VASTA executes that script (Execution phase). The following subsections provide more details about each of these phases and how VASTA utilizes the aforementioned language and vision components.

\subsubsection{Demonstration} 

\par At the start of the demonstration, VASTA navigates to the home screen of the phone and kills all running app processes to ensure that we have a comparable starting point during the execution later. Then, it uses the Android debugging bridge (ADB) connection to capture the current state of the screen (screenshot) at each interaction as well as the type of the user touch events, including taps, long taps, and swipes. For taps and long taps, VASTA logs the coordinates of the click, as well as the duration of it. For swipes, it logs the coordinates of the first and last touch, as well as the duration of the swipe. 





\subsubsection{Learning}

\par VASTA uses a sequence of device screenshots and <X, Y> coordinates of user interaction gathered from the user's demonstration to create an automation script for task execution. Our learning step uses computer vision to detect visual and language information for the UI elements that users interacted with during their demonstration. 


\begin{figure}[h!]
\centering
\includegraphics[width=0.31\textwidth]{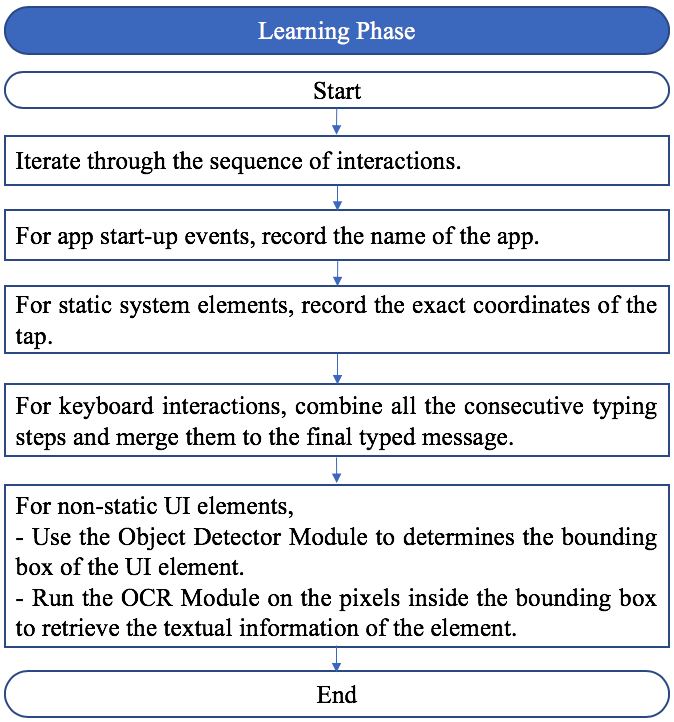}
\caption{The learning phase of VASTA} 
\label{learning}
\end{figure}

\par For every user input event, VASTA uses device screenshots and user traces to determine what type of UI element was interacted with. There are four major types of UI elements a user can interact with: (1) static UI elements: system-level elements, such as the "home" button and the menu drop-down area at the top of the device, whose look and positioning remain the same regardless of the application, (2) app start-up buttons, (3) the keyboard,  and (4) non-static UI elements: elements which fall into neither of the previous three categories. Each of these four types of UI elements is processed differently by VASTA.

For app start-up events, VASTA simply records the name of the app which was launched. For static system level elements, VASTA records the exact coordinates of the tap. For keyboard interactions (which VASTA recognizes by detecting a keyboard on the screen using template matching), VASTA combines all the consecutive typing steps and merges them into on step which includes the final typed message.

The last category, non-static UI elements, is the only type with elements that are susceptible to positional and visual changes during the execution phase, and VASTA cannot predict their new positions during the learning phase. For this reason, VASTA must extract visual and language information for non-static elements during the learning phase as a reference, allowing the user's input to be executed successfully despite these changes. VASTA extracts visual information in the form of bounding box localization. It uses the RetinaNet Object Detector module to extract a rectangular bounding box containing just the UI element in question from the whole device screenshot, selected by using the most probable bounding box that contains the user's click coordinates. VASTA uses this extracted image as a reference to find the element again during the execution phase even if there are translations in its location. VASTA also runs OCR on this extracted image to obtain the text within the UI element, if any. This text is very useful to determine if two elements are one and the same in case of visual changes in the appearance (e.g., the texts below an empty and a full recycle bin are the same, while the icons are different). 

Finally, VASTA saves the extracted visual and language information alongside the user's input into the task automation script. Figure \ref{learning} summarizes the steps of the learning phase.

\subsubsection{Execution}

\par The user can ask VASTA to execute any previously demonstrated tasks saved on their device. VASTA directly executes each ADB command one by one. In the case of app startup events, static system elements, and keyboard inputs, VASTA executes the ADB commands without any modification. In the case of non-static UI elements, the automation script contains extra information with the command (the pixel-representation of the target UI element during the demonstration, the bounding box of the element, and the text within the element). VASTA uses this extra information to determine if the command must be modified. We call this process the element detection procedure which consists of three steps. 



\par I) In the first step, VASTA takes a screenshot of the current screen of the device and searches for the target element in the exact same location as the demonstration using the Template Matching Module. If the template matching succeeds, the ADB command will be executed without change. Otherwise, VASTA goes to the second step in which it checks for positional changes. 

II) In the second step, VASTA uses the Template Matching Module to match the image of the selected UI element during the demonstration with all possible locations in the current screenshot of the device. Template matching uses a sliding window approach to compare a template image across a larger search image and calculates a score at every sliding position. Hence, VASTA can locate the element it is searching for even if there are positional translations of its location. The reason for this design decision is the fast speed and high accuracy of template matching compared to object detection or OCR. In most cases, VASTA can find the target UI element in the first two steps using template matching. We note that template matching succeeds only if there is an identical match and hence, would not make a mistake in the case of templates that are visually similar but not exact, such as mp3 and mp4 icons. If VASTA cannot find the desired UI element in the first two steps, then it is likely the element itself has undergone visual change (e.g., due to app updates), and VASTA goes to the next (third) step and attempts to use language information to identify the correct element to interact with. If the element VASTA is searching for does not have any language information, the execution process stops, as VASTA does not have any more information to locate the correct UI element and fails.

III) In the third step, VASTA attempts to use any textual information within the UI elements to find the target UI element. VASTA uses the RetinaNet Object Detector Module to find all the elements on the current execution screen. After finding all elements on the current execution screen, VASTA uses OCR to retrieve textual information for all found elements. VASTA uses the Levenshtein distance score \cite{yujian2007normalized} to compare the language information obtained during the learning phase with the language information obtained from each element. The Levenshtein distance score is a string metric for measuring the difference between two sequences. Using this language score, VASTA re-ranks the given object detection proposals and returns the top matching element if the distance is lower than a certain threshold\footnote{We consider a Levenshtein distance score of 0.8 as valid, where the score of 1 means denotes a perfect string match. This threshold accounts for minor changes in the string (e.g., caused by imperfect bounding box detection of the element).}. Note that during the execution phase, the object detector is only used to help OCR determining the textual information of individual UI elements. In other words, no pixel-wise similarities are calculated after the template matching steps (first two steps) and VASTA only relies on the textual info of the UI element to detect it. This pipeline is chosen due to its significantly better speed compared to running OCR on the entire image and then grouping the words.


\par After locating the target UI element, VASTA then performs the user's input at the new location. For taps or long taps, the user input will be executed in exactly the same way at the new location. For the user's demonstrated swipes and scrolls that were followed by an interaction with a non-static UI element, during execution, VASTA first ignores the swiping input and searches for the target element (the non-static UI element) in the current device screen using the element detection procedure (the 3 steps explained above). If VASTA cannot find the element in the current screen, it swipes the screen in the same direction as demonstrated by the user\footnote{The duration and distance of the swipe by VASTA during execution are fixed and chosen based on the device's specification to ensure that each swipe moves 90\% of the screen's width or length.} and again, searches for the target UI element. VASTA continues this process until it finds the element. With this procedure, we ensure that even if the target element moves up/down/left/right in a list of UI elements after the demonstration phase, VASTA still finds it. 

\par
\begin{figure}[h!]
\centering
\includegraphics[width=0.31\textwidth]{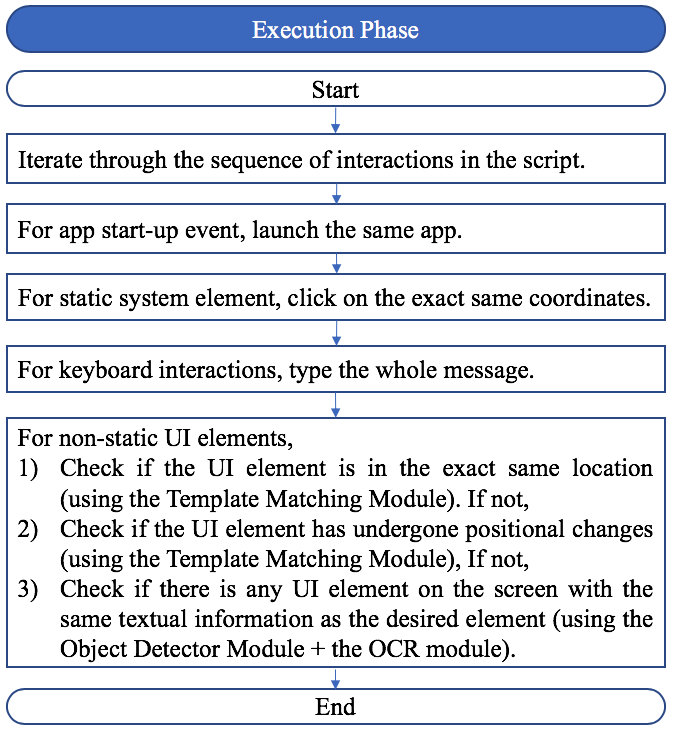}
\caption{The execution phase of VASTA} 
\label{fig:executionphase}
\end{figure}

 After running all instructions on the execution script, VASTA then ends the execution process and goes back to being on standby. Figure \ref{fig:executionphase} illustrates the flowchart process of VASTA's execution phase.

\section{Experiments} \label{sec:userstudy}

\subsection{Object Detector} \label{subsec:expObjectDetector}
In order to develop an object detector to detect UI elements on the screen, first, we investigate two state-of-the-art object detector networks, namely RetinaNet (with a ResNet-50-FPN backbone) \cite{lin2018focal} and YOLOv3 \cite{redmon2018yolov3}. Other than the ability to accurately determine the bounding boxes of UI elements, the inference time of the object detector module was essential to us as VASTA uses this module during both the learning and the execution phases. 

To train the object detector models, we used the RICO dataset \cite{deka2017rico}, which consists of over 72k unique UI screens across 9.7k free Android apps. Each UI screen has its own detailed view hierarchy, which contains accessibility information about every UI element within the screen. Of these elements, we only work with those with the following properties: "clickable", "visible", "visible-to-user", and "enabled". We divided the RICO dataset randomly into three partitions of approximately 52k, 10k, and 10k images for training, hyper-parameter optimization, and testing, respectively. \footnote{We trained this network on 8 Tesla M40 GPUs with a batch size of 16 images per GPU. The network was initialized with ImageNet weights, and we fine-tuned the network on the 52k RICO image subset until the loss settled at 30 epochs. We use an Adam Optimizer with a learning rate of 1e-5 and default training parameters beta 1 = 0.9 and beta 2 = 0.999.} As shown in Table \ref{table:yoloVSretina}, between the two models, RetinaNet provided a considerably higher Average Precision (AP) scores on the test set in different Intersection Over Unions (IOUs). Although YOLOv3 runs more than 2x faster, RetinaNet runs fast enough to provide a smooth execution phase. Hence, we chose this model as our object detector.

\begin{table}[!h]
        \centering
	\caption{Performance of Object Detection Networks on the RICO dataset}
	\begin{tabular} {c | c | c |c | c}
		\hline
		\small Network & \small  mAP@[.5:.95])   & \small AP@.50 & \small AP@.75 & \small Inference Time  \\
		\hline
		\small RetinaNet  & \small 58.49        & \small  85.04     & \small 60.46   & \small   52 ms/img       \\
		\small YOLOv3 & \small 27.99     & \small 73.58      & \small 17.72   & \small 23 ms/img      
	\end{tabular}
	\label{table:yoloVSretina}
\end{table}

\subsection{User Study - Language Component}
In order to evaluate the effectiveness of the natural language understanding modules, we conducted a user study in which participants provided utterances for different tasks.

\subsubsection{Participants and Tasks}
10 participants (7 males), all proficient in English, were recruited for this study. We asked each participant to imagine a scenario in which they want a personal assistant to carry out one or more tasks. For each of these tasks, we asked them to write down variations of utterances that they might use to talk to the assistant. For example, participants were told ``Say that you want your personal assistant to carry out the task of finding the price of Uber pool to the train station. Write down at least 5 different ways in which you might tell the assistant to do this task.'' We collected a total of 100 utterances belonging to 10 different tasks (i.e., 10 utterances per task), which are listed in Table \ref{table:nlp-user-study-tasks}. We call this the \textit{Original Dataset}. Then we synthetically introduce changes in the utterance parameters so that no two utterances have the same parameters. For example, the utterance ``What's the price of Uber pool to the Second street?'' is manually changed to ``What's the price of Uber X to the train station?''. We call this the \textit{Synthetic Dataset}.

\begin{table*}[t!]
        \centering
	\caption{List of tasks in the user study of the language component, along with a sample utterance provided by a participant for each task.}

	\begin{tabular}{p{8.5cm} | p{8.5cm}}
	\small
		\textbf{Task} & \textbf{\small Sample utterance} \\
		\hline
\small
		Tell the assistant to find nearest Italian restaurants. & \small Give me the closest Italian restaurants.\\
		\hline
\small
		Tell the assistant to message the `myteam' Slack channel that you'll be late.  & \small Tell myteam channel in Slack that I'll be late.\\
		\hline
\small
		Tell the assistant to take a selfie and send it to your mom on Whatsapp. & \small Can you take a selfie and share it with my mom on Whatsapp?\\
		\hline
\small
		Tell the assistant to order a small pepperoni pizza from Domino's. & \small Get me a small pepperoni pizza from Domino's.\\
		\hline
\small
		Tell the assistant to show you your grade in the course CS101. & \small I want to see my grade for CSC101.\\
		\hline
\small
		Tell the assistant to search Netflix for Al Pacino movies. & \small What's on Netflix from Al Pacino?\\
		\hline
\small
\small
		Tell the assistant to show the statistics of the basketball player, Lebron James. & \small Please tell me the stats of Lebron James.\\
		\hline
\small
		Tell the assistant to find the next Barcelona game using the Yahoo Sports app. &  \small Open Yahoo Sports app and find the next match of Barcelona.\\
		\hline
\small
		Tell the assistant to find out the price of the tickets to the Metallica concert. & \small Show me the ticket price for the Metallica concert.\\
		\hline
\small
		Tell the assistant to give you the cost of Uber pool to go to the train station. & \small Can you find out how much Uber pool costs to take me to the train station?
	\end{tabular}
	\label{table:nlp-user-study-tasks}
\end{table*}

\subsubsection {Results}
\begin{figure}[b!]
	 \centering
		\includegraphics[width=0.95\linewidth]{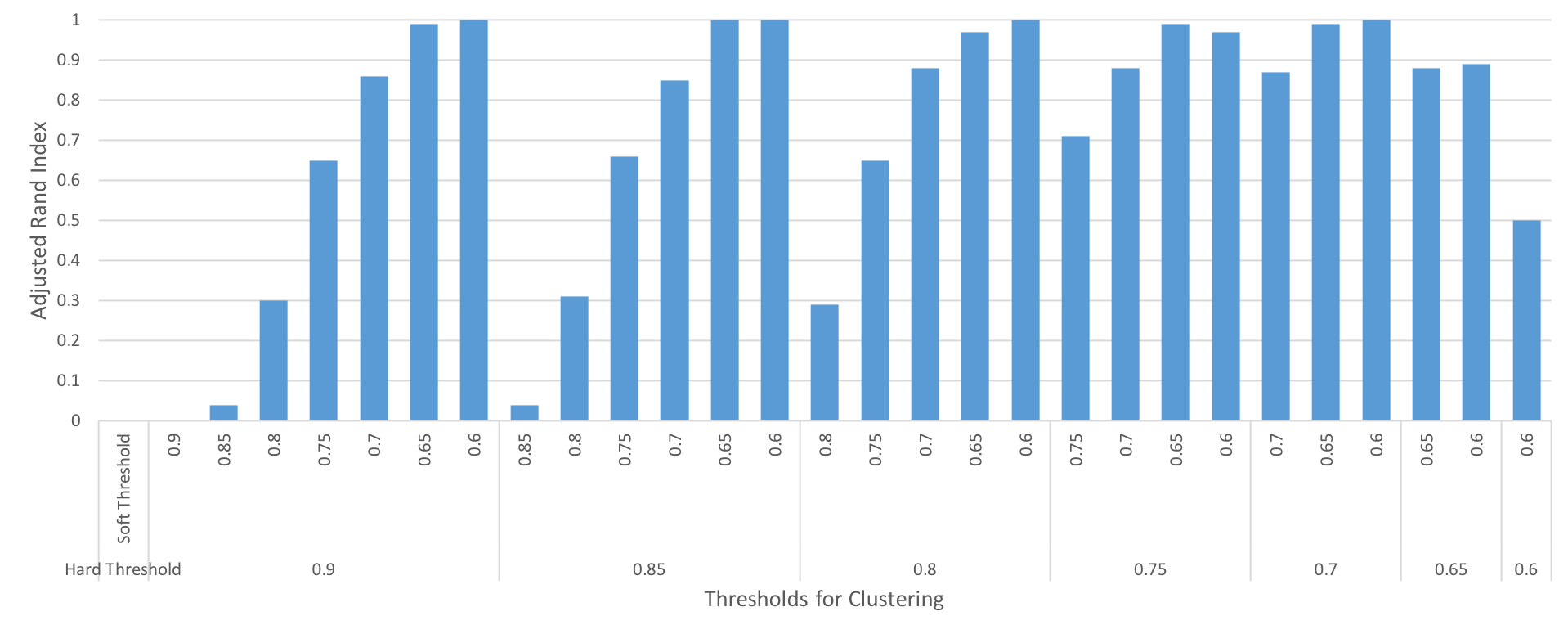}
	\caption{Accuracy of Utterance Clustering Module with varying hard and soft threshold values.}

	\label{fig:clustering-accuracy}
\end{figure}

\begin{figure}[b!]
	\centering
		\includegraphics[width=0.95\linewidth]{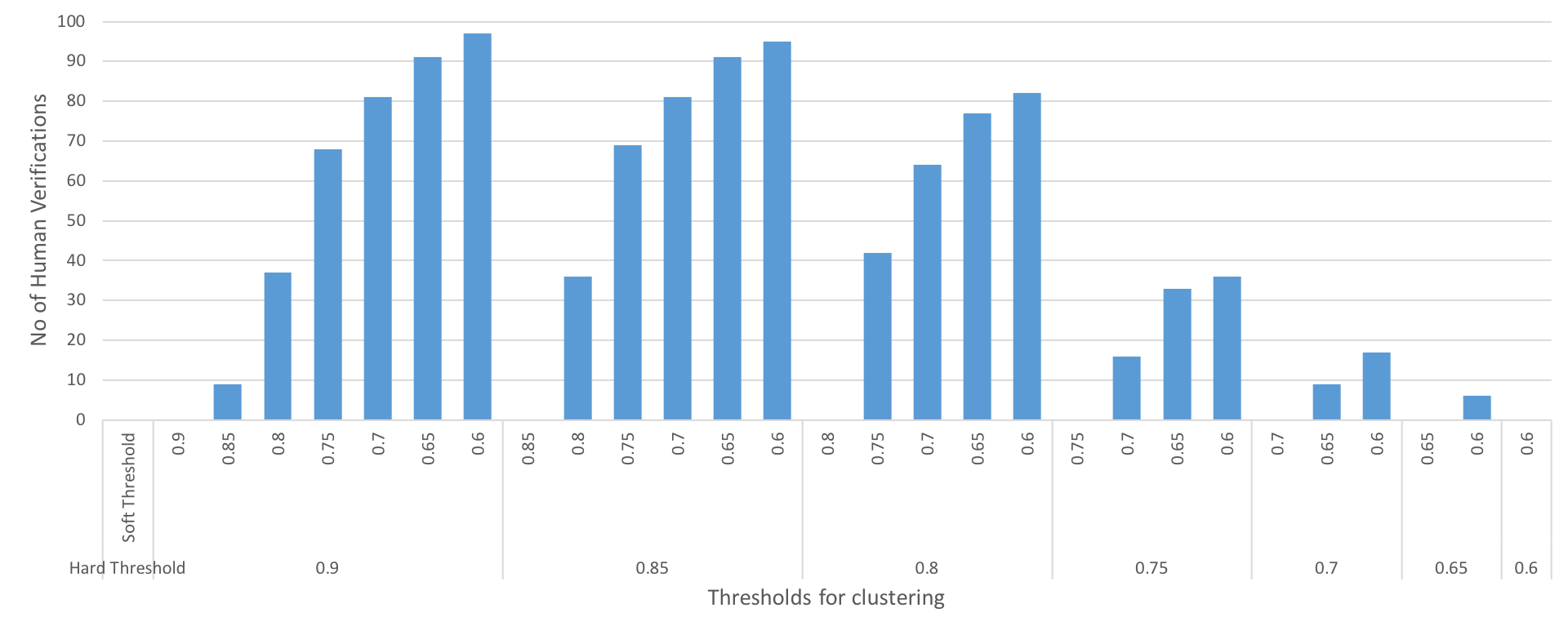}
	\caption{Number of user verifications in Utterance Clustering Module with varying hard and soft threshold values.}

	\label{fig:clustering-verifications}
\end{figure}

We randomized the sequence of utterances in a dataset and input them one by one into the Utterance Clustering Module, and if applicable, into the Parameter Prediction Module. We repeated this process 10 times each for both the original and synthetic datasets. The accuracy of clustering is measured using the adjusted Rand index \cite{rand1971objective}. It calculates the percentage of agreement of pairs of elements, i.e., whether they belong to the same cluster or not, between the predicted and ground truth clusters, corrected for chance. We varied the hard and soft thresholds in the range [0.6, 0.9], for which we report the clustering accuracy results in Figure \ref{fig:clustering-accuracy}. We also report the corresponding number of user verifications in Figure \ref{fig:clustering-verifications}. For brevity, we only report these results on the \textit{Synthetic Dataset} but find the trends in the \textit{Original Dataset} to be similar. Overall we find that using a hard threshold of 0.7 and a soft threshold of 0.6 lead to perfect clustering (i.e., adjusted Rand index of 1.0) while ensuring minimal user verifications (i.e., 17 verifications out of a total of 100 utterances). Since our clustering is incremental, each incoming utterance may be assigned to an existing or a new cluster. Hence, our evaluation method accounts for errors not just in assignment to incorrect clusters, but also for errors in creating fewer or more clusters than desired.


In order to evaluate the Parameter Prediction Module in isolation, we conducted another experiment in which every pair of utterances belonging to the same task was input to the Parameter Prediction Module with the first utterance acting as the canonical utterance of a cluster and the second utterance acting as an input utterance newly matched with the same cluster. Then we calculate the average accuracy of an exact match, which is a stringent measure that only rewards exact parameter string matches. For example, in the utterance ``Find an Italian restaurant near me'', if `an Italian' is predicted as the parameter instead of `Italian', this will be treated as incorrect. We also calculate a comparatively lenient set of measures at the word level -- precision, recall and F1-measure. These results are reported in Table \ref{table:ppm-results}. The F1-measure for word match is 82\%  for the \textit{Original Dataset} and 65\% for the \textit{Synthetic Dataset}.

\begin{table}[!h]
        \centering
	\caption{Performance of Parameter Prediction Module}
	\begin{tabular} {c | c | ccc}
		& \small Exact Match & \multicolumn{3}{c}{\small Word Match}  \\
		\hline
		\small Datasets & \small Accuracy    & \small Precision & \small Recall & \small F1 \\
		\hline
		\small Original  & \small 69\%        & \small 69\%      & \small 99\%   & \small 82\%       \\
		\small Synthetic & \small 40\%        & \small 53\%      & \small 85\%   & \small 65\%      
	\end{tabular}
	\label{table:ppm-results}
\end{table}

\subsection{User Study - End to End}
\subsubsection {Participants and Tasks}
\par 10 participants (6 males) aged from 21 to 45 (mean $= 31.5 \pm 6.3$) were recruited for this study. They were required to be fluent in English and active smartphone users. 

\par To determine the tasks for the user study, we first conducted a survey, in which participants were asked to provide a list of their common repetitive smartphone tasks, and we chose seven tasks (see Figure \ref{userstudyfails}) from this list (one for tutorial only). None of the selected tasks can be currently executed by smartphone intelligent assistants such as Google, Siri, or Bixby. The motivation behind choosing each task is also provided in Figure \ref{userstudyfails}. 

At the start of the study session, the researcher described the tasks and played a video for the participant to show him/her how each task needed to be done. For each task, participants were instructed to ask VASTA (with no previously learned task) to carry out that task. They were free to choose the utterance they use for each task. Since VASTA had no previously learned task, the participant needed to demonstrate the task. When the participant finished the demonstration of all tasks, he/she was instructed to ask VASTA again to carry out each task. Again, participants were not limited to use any specific utterance. However, they were instructed not to use the exact utterance they used the first time so that the performance of VASTA's Utterance Clustering Module can be evaluated. Also, for three out of six tasks (tasks 1, 3, and 5), participants were instructed to change the parameter(s) of the utterance as well so that the performance of the Parameter Prediction Module can be assessed.

We designed VASTA to be robust to the UI's visual changes (i.e., the UI's appearance due to updates) and positional changes (i.e., movement of the location of UI elements). In order to evaluate VASTA's robustness to positional changes, for three out of the six tasks (tasks 2, 3, and 4), we made sure that the locations of some UI elements are different in the demonstration phase and the execution phase. To evaluate the effect of visual changes due to a version update, we synthesized an app (task 6) specifically for this experiment. \footnote{To simulate the possible UI changes of an app after a version-update, we develop two versions of a school-grades app with significant visual differences in the UI, but same textual information on the elements.} More details about each task are provided in Figure \ref{userstudyfails}.

\subsubsection {Results}

\begin{figure*}[!t]
 \centering
    \includegraphics[width=\linewidth]{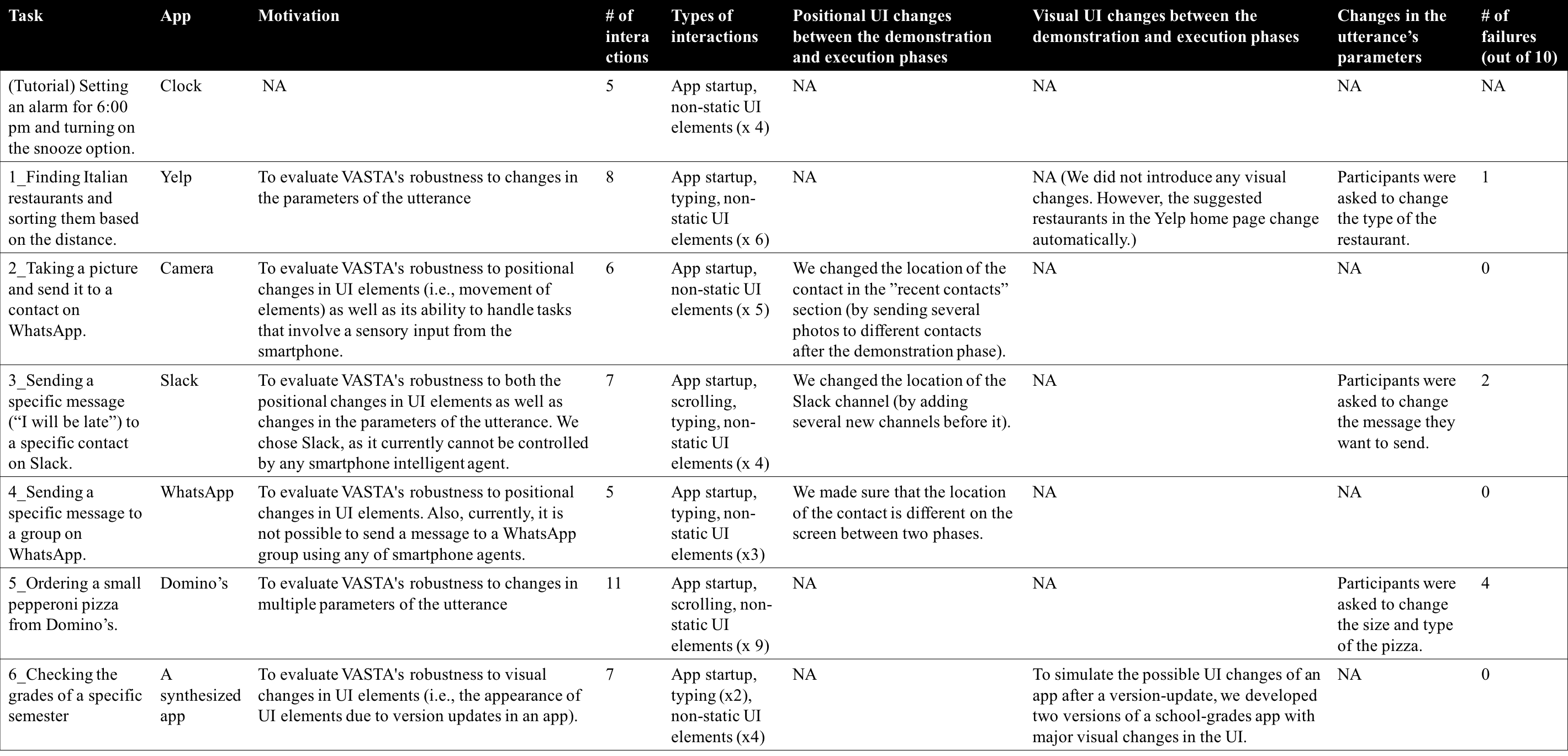}
\caption{The details of the tasks of the end-to-end user study. The last column presents the results.}
  \centering
\label{userstudyfails}
\end{figure*}

\par Overall, the execution of 53 out of 60 scripts (10 participants, 6 tasks each) that VASTA created from the study were successful. Figure \ref{userstudyfails} shows the number of failures for each task separately. Here, we provide details on how these errors break down across various modules of VASTA.

\begin{figure}[!h]
\centering
    \includegraphics[width=0.88\linewidth]{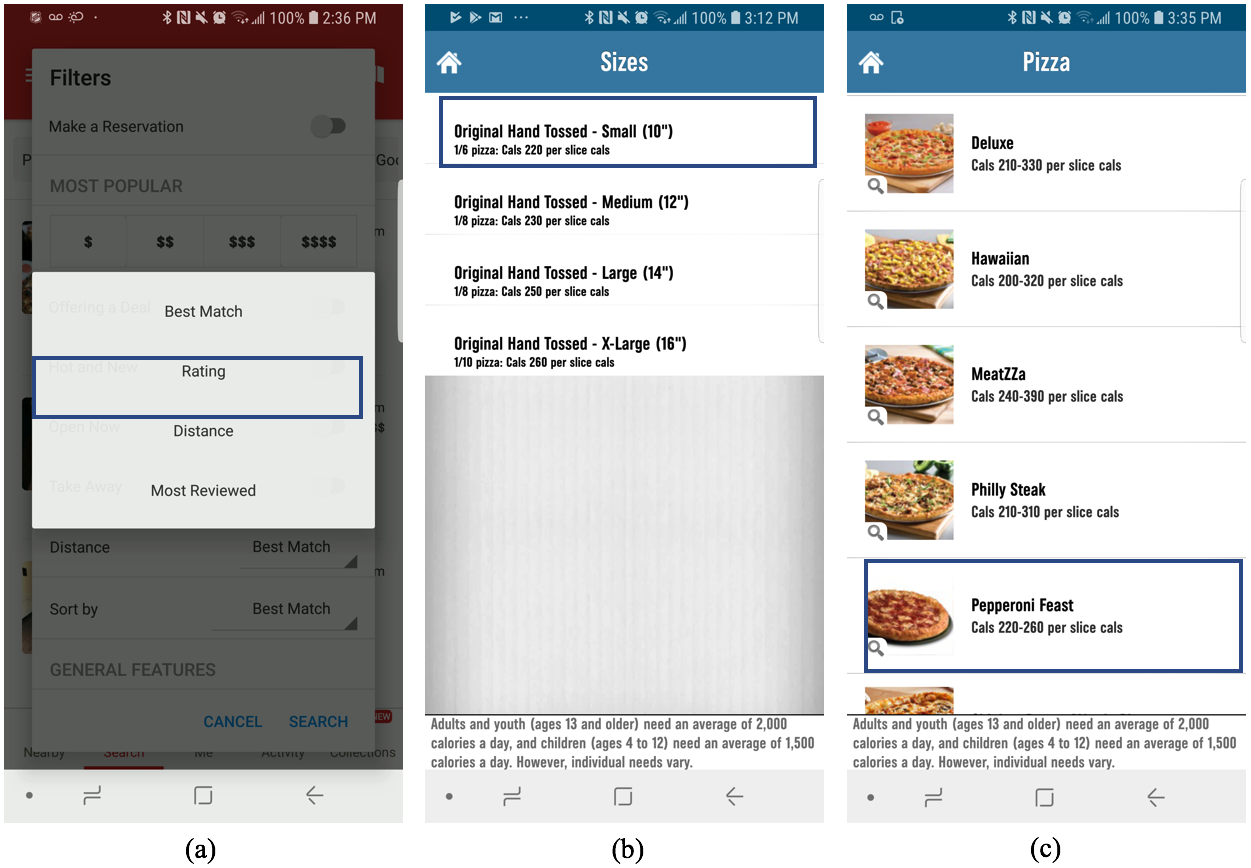}
      \caption{Three examples of failure cases of the user study. In (a), the user clicked on the "Distance" option, but the object detection network mistakenly detected that "Rating" is the desired UI element. In (b) and (c), VASTA failed to generalize the utterance parameters (pizza size and type) due to the amount of unnecessary text in the detected bounding boxes.}
  \centering
\label{mistakes}
\end{figure}

\textbf{Object Detector + OCR (Learning Phase)} In 59 out of 60 scripts, VASTA was able to find all correct UI elements that the user interacted with. The only mistake happened in the Yelp task when one user was supposed to click on "Distance" under the "Sort by" menu (see \ref{mistakes}(a)). Since the coordinates of the user's click were also close to the word "Rating", the object detection network mistakenly detected that "Rating" is the desired UI element in this step (as VASTA uses the most probable bounding box that contains the user's click coordinates). Hence, VASTA mistakenly chose the "Rating" option as the user's selection for that step.

\textbf{Utterance Clustering Module} VASTA correctly clustered all 60 utterances into 6 tasks, and hence, always started executing the correct task.

\textbf{Parameter Prediction Module} For 47 out of 60 utterances (78\%), VASTA predicted the exact correct parameters. Out of the 13 imperfect prediction cases, 6 still led to correct executions. 
\begin{itemize}
    \item For 2 utterances (in the pizza ordering task), the use of Levenshtein distance score helped VASTA detect correct elements.
    \item For 4 utterances (in the Slack task), more than 90\% of the parameter's characters were correct (e.g., VASTA typed ``that I'm working from home today" instead of ``I'm working from home today").
\end{itemize}
Hence, for 53 out of 60 utterances (88\%), the output of this module led to correct executions. 

For 7 utterances (12\%), the incorrect predictions caused failure in the execution phase.
\begin{itemize}
    \item For 4 utterances (in the pizza ordering task) which contained 2 parameters each (i.e., pizza size and type), our algorithm either swapped the parameters or predicted both as part of a single parameter. This happened because both parameter's tokens have very similar nodes in the dependency tree (i.e., both are adjectives and are attached to the noun `pizza').
    \item For 3 utterances (in the Slack task), our algorithm either failed to predict all the correct tokens or predicted many additional incorrect tokens in the parameter. This happened in cases when the utterance was long and/or contained at least 2 subordinate clauses (e.g., `please use slack to send a message to my team that i am working from home').
\end{itemize}

\textbf{Template Matching Module} This module worked successfully in 100\% of the cases. 

\textbf{Object Detection + OCR (Execution Phase)} In all the cases where every other module worked without any error, these modules also worked perfectly. However, in the four complete failure cases of the Parameter Prediction Module in the pizza ordering task (mentioned above), the imperfect performance of the object detection and OCR components made it hard for VASTA to correct those mistakes. Figures \ref{mistakes}(b) and (c) show two examples of these imprecise bounding boxes. In Figure \ref{mistakes}(b)), the mistake happened because instead of clicking on the text, the user clicked on the white area on the right side of the text, and hence the object detection system recognized the entire button as the UI element (containing other information about that size, such as the number of slices, the diameter in inches and the number of calories ). In Figure \ref{mistakes}(c), VASTA made a similar mistake in choosing the type of pizza. Again, unnecessary information is present on the element, such as the number of calories. Also, the OCR algorithm falsely detects some characters in the picture of the pizza, which happened to be different for each pizza type. 

The 13 cases of imperfect parameter prediction show the risk of generalizing utterance parameters without validating it with the user. Notably, this risk is higher when the parameters are written on UI elements and not typed directly by the user (e.g., the four complete failure cases of the pizza ordering task).

\par
\begin{figure}[!h]
  \centering
    \includegraphics[width=0.9\linewidth]{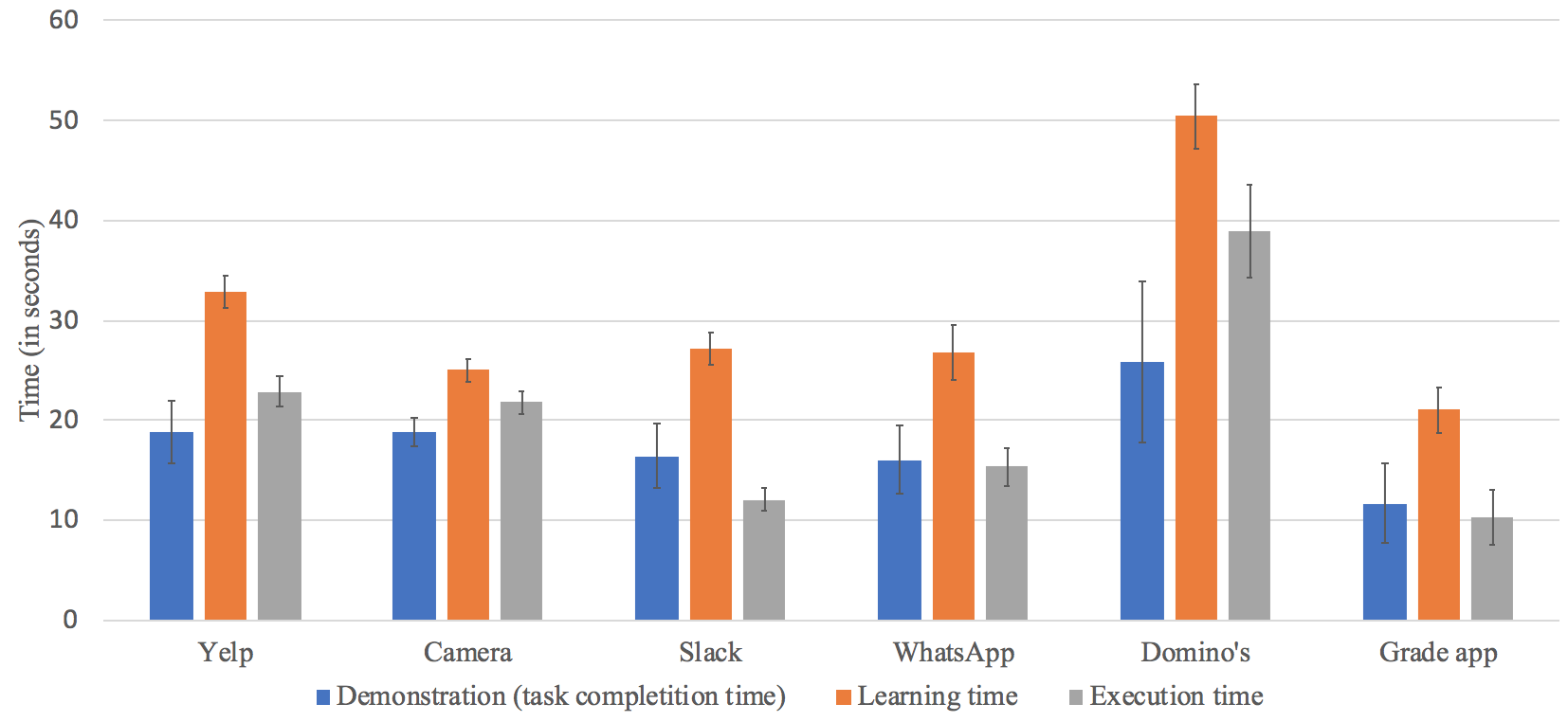}
      \caption{The average demonstration, learning, and execution period for each task.}

\label{userstudytimes}
\end{figure}

 Another critical performance measure for VASTA is the system's execution speed compared to the user's demonstration speed for the same task. Figure \ref{userstudytimes} shows the average demonstration, learning, and execution period for each task. For 5 out of 6 tasks (except task 5), there is no significant difference (Wilcoxon signed rank test) between the demonstration and execution durations. For the Yelp and Camera tasks, the average execution duration was slightly longer than the demonstration duration, while for the Slack, WhatsApp and the grades app, the execution phase was shorter. The reason for this difference is the presence of a relatively long typing step in these three tasks. While the user uses the smartphone's keyboard to enter a message letter by letter, VASTA types the entire message in one step which saves time. 

In the case of the Domino's app, the execution time is about 1.5 times longer than the demonstration duration. The reason for this increase is the limited speed of the OCR algorithm in the swiping-down step. To reach the target pizza type icon, VASTA needs to swipe down four times, and after each swipe, VASTA searches for the target UI element on the screen which includes first searching for template matches and then (since no exact match can be found), detecting all UI elements and running OCR on them one by one. With the OCR implementation that we use, this step takes considerably longer compared to the other two steps (template matching and object detection). For the Domino's task, on average across participants, the OCR step was responsible for 59\% of the execution duration. In future versions of VASTA, we plan to investigate the possibility of running a lighter version of OCR on the entire screen and then using object detection bounding boxes to assign characters to different elements.  

\par After the end of the experiment, participants answered several questions about their experience. Participants were asked to rate their agreement with statements related to their experience interacting with VASTA on a 5-point Likert scale from "Strongly Disagree" to "Strongly Agree." Table \ref{tab:survey} depicts the average score for all the statements in the survey. We also asked participants if they have any suggestion for enhancement of VASTA as a task automation agent and summarize the collected suggestions in the future work section of this paper. 
\begin{table}[h!]
\begin{center}
\caption{Average scores on usability questions from the
post- questionnaire on a 5-point scale (1- Strongly disagree, 2-Disagree, 3-Neutral, 4-Agree, 5-Strongly agree).} \label{tab:survey}
\begin{tabular}{ l c c}
\hline
\multirow{ 2}{*}{\small Statement} &\multicolumn{2}{c}{\small Score} \\ 
& \small Mean & \small STD \\ \hline 
\small I find VASTA useful in helping me creating automation. & \small 4.5 & \small 0.5 \\
\small I feel that VASTA is safe to use. & \small 4.2 & \small 0.8 \\
\small I would use VASTA to automate my tasks. & \small 4.2 & \small 0.8 \\ \hline 

\end{tabular}
\end{center}
\end{table}

\section{Discussion} \label{sec:discussion}
We designed VASTA to address some of the main challenges in the development of a smartphone task automation system using PBD: robust recognition of the user utterance and generalization of the automation, usability with any arbitrary third-party app, robustness to positional and visual changes in the UI, and changes in the parameters of the utterance. Among the previously reported PBD-based smartphone task automation systems, \textsc{Sugilite} is the only one which partly addresses some of these issues. However, its reliance on the XML structures of apps, provided by the Android accessibility API, yield several shortcomings which we will discuss here.

Despite the success of markup based PBD methods, such as those leveraging XML or HTML, these methods will eventually fail in the face of ambiguity or when markup is not available. For example, when there is ambiguity, SUGILITE prompts the user to clarify which UI object in the view hierarchy they wished to interact with~\footnote{What appears as a single UI element on a page, might be composed of multiple views}. This is challenging for end users who are not developers. Also, markup can often be unavailable, such as Android applications which are written as a web view embedded inside a native wrapper. The markup language leveraged by SUGILITE (Android XML) and other applications would no longer be available in the web view. And further, even if the markup in the web view (HTML) was usable by an application like SUGILITE, it could further embed other content that may not be readable, such as WebGL elements. In the absence of the complex engineering required to fetch, parse, and interpret all of these various markup languages, vision-guided techniques can be used to supplement traditional methods. Furthermore, because VASTA has no dependency on the UIs underlying markup language, it can be applied to systems where none is available, such as those UIs rendered with low-level graphics libraries.

We also tried to compare the performance of VASTA's language component with SUGILITE. The main author provided helpful assistance in getting started with SUGILITE, but we ultimately failed to do a side-by-side comparison of VASTA and SUGILITE due to technical difficulties. Despite our best efforts, the latest implementation of SUGILITE did not work in any scenario of our user study due to a variety of reasons (e.g., lack of support for scrolling, bugs in text entry recording mechanism, etc.). These issues were confirmed by the author.

\subsection{Future work}
\subsubsection{Semantic labeling of UI elements}
\par Currently, VASTA only records a feature representation of UI elements to track and find these elements later in the execution step. However, assigning a semantic label (e.g., a "sign-in" button or a "send" icon) to each UI element using an image classification network, such as the one described in \cite{liu2018learning}, can help the accuracy of the system in detecting the same element during execution. Also, it can be used to prepare a storyline for each task which can be potentially helpful in generalizing a task which was done in an app to be executed using a different app (e.g., sharing a link on Linkedin during the demonstration step and executing the script using Facebook).
\subsubsection{Using multiple apps}
\par VASTA can currently automate a task including multiple apps. However, we do not currently support transferring data from one app to another (e.g., finding the arrival time of the next bus and sending it to a contact), which would require a mechanism to allow users to indicate important data within an application. This capability was suggested by most of the participants in our user study. 

\subsubsection{Combining computer vision with XML data}
\par Using object detection and computer vision to detect and record UI elements instead of XML data fixes many of the disadvantages of PBD systems based on accessibility APIs. However, as can be seen in the user study results, the object detection network makes mistakes which can cause the automation to fail. An interesting approach to minimize the number of failures is to develop a task automation agent by leveraging both XML data (whenever they are available), such as \cite{SUGILITE}, and computer vision techniques, such as the ones used in this paper. This can also help to overcome a few limitations of our language component, two of which are mentioned here. First, VASTA may not be able to distinguish between tasks with similar command structures but different parameter values (e.g., ``Get me tickets to Metallica'' in a concert app and ``Get me tickets to Avengers" in a movie app). Second, given a user utterance, it cannot handle discrepancies between the utterance and the demonstration. For example, if the user utters ``Find the nearest Italian restaurants", but searches for Korean restaurants during the demonstration, it will not alert the user about the discrepancy. Instead it will not find any slots and will incorrectly assume that the task corresponds to finding Korean restaurants. In the future we aim to overcome these limitations by leveraging knowledge of all possible parameter values in a task using a combination of XML information and computer vision techniques.

\section{Conclusions} \label{sec:conclusions}
We described VASTA, a PBD system for smartphone task automation that leverages recent advances in computer vision and natural language processing to effectively recognize automation tasks, predict automation task parameters, and detect positional and visual changes in UI elements during a demonstration. To the best of our knowledge, VASTA is the first system to leverage computer vision techniques for smartphone task automation. With the help of user studies, we demonstrated VASTA's adaptability to many changes in an application's UI elements as well as its generalizability to users' utterance variations. This system is potentially applicable to automation across different operating systems and platforms.

%
%
%
%
%

\bibliographystyle{SIGCHI-Reference-Format}
\bibliography{paper}

\end{document}